\documentclass[conference]{IEEEtran}
\newif\ifblinded\blindedfalse

\usepackage{cite}
\usepackage{amsmath,amssymb,amsfonts}
\usepackage{algorithmic}
\usepackage{graphicx}
\usepackage{array,booktabs}
\usepackage{subcaption}
\usepackage{comment}
\usepackage{url}
\usepackage[urlcolor=blue]{hyperref}
\usepackage{xcolor}
\usepackage{siunitx}
\usepackage{pifont}

\newcolumntype{P}[1]{>{\centering\arraybackslash}m{#1}}

\definecolor{colgood}{rgb}{0.0,0.45,0.0}
\definecolor{colbad}{rgb}{0.6,0.0,0.0}
\newcommand{\cmark}{\textcolor{colgood}{\ding{51}}}
\newcommand{\xmark}{\textcolor{colbad}{\ding{55}}}
\newcommand{\R}{\mathbb{R}}

\newcommand{\tm}{\ifblinded\else ™\fi}

\newcommand{\todo}[1]{{\color{red} TODO: #1}}
\newcommand{\fref}[1]{Fig.~\ref{#1}}
\newcommand{\tref}[1]{Tab.~\ref{#1}}
\newcommand{\sref}[1]{Sec.~\ref{#1}}
\newcommand{\eref}[1]{Eq.~\ref{#1}}

\newcommand{\sdv}{SDV}
\newcommand{\bseg}{BSEG}

\begin{document}

\title{
    Arithmetic Packing on Wide Integer Datapaths in DSP Primitives of Modern FPGA Devices
}

\author{%
\ifblinded
    \IEEEauthorblockN{-- Blinded for Review --}
\else
    \IEEEauthorblockN{%
        Titus Bornträger\IEEEauthorrefmark{1}, Shane Fleming\IEEEauthorrefmark{2}, %
        Philipp Holzinger\IEEEauthorrefmark{1}, Dietmar Fey\IEEEauthorrefmark{1}, %
        Michaela Blott\IEEEauthorrefmark{3}, Thomas B. Preußer\IEEEauthorrefmark{4}%
    }
    \IEEEauthorblockA{%
        \IEEEauthorrefmark{1}%
        \textit{Friedrich-Alexander Universität Erlangen-Nürnberg}, Erlangen, Germany, \textsl{firstname.lastname}@fau.de%
    }
    \IEEEauthorblockA{%
        \IEEEauthorrefmark{2}%
        \textit{AMD Research and Advanced Development}, Swansea, UK, shane.fleming@amd.com%
    }
    \IEEEauthorblockA{%
        \IEEEauthorrefmark{3}%
        \textit{AMD Research and Advanced Development}, Dublin, Ireland, michaela.blott@amd.com%
    }
    \IEEEauthorblockA{%
        \IEEEauthorrefmark{4}%
        \textit{AMD Research and Advanced Development}, Dresden, Germany, thomas.preusser@amd.com%
    }
\fi
}

\maketitle

\begin{abstract}
Deep Neural Networks increasingly employ low-precision quantization to reduce computational requirements. While FPGAs are well suited for workloads with heterogeneous precisions, their dedicated digital signal processing (DSP) slices only feature fixed-width datapaths that are significantly underutilized by low-bitwidth arithmetic. While previous approaches have already introduced the packing of multiple values onto the same wide DSP datapath, they either only support specific fixed bitwidths or are wasteful regarding the use of additional support logic external to the DSP. This paper proposes an efficient method to dynamically pack multiple (un-)signed inputs with arbitrary bitwidths into a wide multiplier path by leveraging the DSP's internal pre-adder. Building on this, we present two distinct architectures, one optimized for matrix-vector multiplications and the other for convolutions. Our implementations are integrated into AMD’s FINN framework. With these optimizations, we reduce the LUT utilization by 21\% and increase the FPS/DSP by 36\% for the UltraNet model compared to the FINN reference.

\end{abstract}

\begin{IEEEkeywords}
AI accelerators, digital signal processors, field-programmable gate arrays, fixed-point arithmetic
\end{IEEEkeywords}
\section{Introduction} \label{sec:intro}
Modern applications, particularly in the field of artificial intelligence, require increasingly higher compute power at a low energy consumption~\cite{thompson2020computational}.
Field-programmable gate arrays (\mbox{FPGAs}) have emerged as indispensable platforms for satisfying this need in the embedded application domain~\cite{boutros2025field}.
However, the available resources, such as lookup tables (LUTs), flip-flops (FFs) and digital signal processing slices (DSPs), have not scaled at the same pace as the computational demand.
Hence, the gap between demand and supply of compute resources on hardware devices is growing.
While LUTs and FFs can be used flexibly with bit-level granularity, the dedicated DSP slices provide only a fixed-width, relatively wide multiply-accumulate datapath.
The naive mapping of low-bitwidth operations to these hardware structures results in a significant and systematic underutilization of the consumed silicon footprint.

One example of such applications that heavily rely on low-bitwidth operations are deep neural networks (DNNs).
Here, optimizations are critically needed to reduce the memory and compute demands of large models such that they can also be deployed on resource-constrained edge devices.
For this purpose, sophisticated quantization techniques are often employed to derive low-precision, fixed-point inference versions from models originally trained using floating-point number formats~\cite{sze2020efficient,umuroglu2017finn,blott2018finn}.
These computations, in turn, also have to be efficiently mapped onto the limited amount of available FPGA device resources.
As a solution, previous research presented the packing of multiple arithmetic operations into the same wide DSP datapath~\cite{fu2016deep,preusser2020vectorization}.
However, these existing approaches have significant limitations.
They either only support a specific fixed bitwidth or are wasteful regarding the use of additional support logic external to the DSP~\cite{fu2016deep,liu2022hikonv,yang2024sa4}.
Therefore, efficient resource utilization, particularly in the context of more complex operations like convolutions, still poses a challenge today.


To solve these challenges, this paper proposes multiple optimizations that significantly improve the operational density of a design (i.e., the number of operations per DSP and cycle).
First, we generalize previous approaches to efficiently pack both signed and unsigned values with arbitrary input widths into a wider DSP datapath, particularly for computations smaller than 8~bits.
This is achieved by initially adapting existing research~\cite{preusser2020vectorization} to determine the interactions between all individual products.
Based on this, we present a more efficient new method that dynamically handles signed values while only using the internal pre-adder common to most modern DSP architectures.
We then further extend these techniques to also calculate packed multiply-accumulate operations that are found in convolutions even more efficiently.
For this purpose, we leverage guard bits as lane offsets to separate the individual products.
With these techniques combined we can minimize the required amount of FPGA resources external to DSPs.


To demonstrate the advantages of our approach in practice, we integrated these optimizations into AMD’s state-of-the-art FINN framework~\cite{umuroglu2017finn, blott2018finn} to generate DNN architectures for efficient low-precision matrix-vector multiplications and convolutions on FPGAs.
These implementations are directly optimized for FINN's standard tensor layout, which specifies how multidimensional tensors are mapped onto 1D streams.
Our contributions are also provided as open source, such that all resulting designs are fully accessible and reproducible.

The paper is organized as follows:~\sref{sec:related} reviews previous related works.~\sref{sec:implementation} details the proposed optimization methods and the corresponding architectures for matrix-vector multiplications and convolutions.
These are then evaluated in~\sref{sec:evaluation}.
Finally,~\sref{sec:conclusion} summarizes the findings and gives an outlook on possible future work.

\begin{table*}
\renewcommand{\arraystretch}{1.25}
\caption{Comparison of key features and capabilities of different DSP packing implementations.}
\label{tab:features}
\centering
\begin{tabular}{>{\bfseries}m{2.6cm}P{1.7cm}P{1.7cm}P{1.7cm}P{1.7cm}P{1.7cm}P{1.7cm}P{1.7cm}}
\toprule
  \textbf{Feature} &
  \textbf{INT8}\newline\llap{\cite{fu2016deep}} &
  \textbf{HiKonv}\newline\llap{\cite{liu2022hikonv}} &
  \textbf{SA4}\newline\llap{\cite{yang2024sa4}} &
  \textbf{Overpacking}\newline\llap{\cite{sommer2022dsp}} &
  \textbf{SSiMD}\newline\llap{\cite{liu2023ssimd}} &
  \textbf{DLPack}\newline\llap{\cite{mohabbati2025dlpack}} &
  \textbf{Proposed Methods} \\
\midrule
Packing Strategy & Single & Dual & Dual & Dual & Dual & Dual & Single \& Dual \\
Precision Support & Fixed (INT8) & Arbitrary & Fixed (INT4) & Arbitrary & Fixed (INT4) & Fixed (INT2) & Arbitrary \\
Target Application & MatVec \& Conv & Conv & Conv & Conv & Conv & Conv & MatVec \& Conv \\
Signed Input Packing & Internal (Trivial) & External & External & External & External (Optimized) & \textcolor{colbad}{Unsigned Values Only} & Internal (DSP Pre-Adder)\\
Accurate Computing & \cmark & \cmark & \cmark & \xmark & \cmark & \cmark & \cmark \\
Open-Source & \xmark & \xmark & \cmark & \xmark & \xmark & \xmark & \cmark \\
\bottomrule
\end{tabular}
\end{table*}

\section{Related Work} \label{sec:related}

Prior research on improving the computational throughput of DSPs has introduced a variety of techniques that differ significantly in their implementation details. To highlight our contributions,~\tref{tab:features} provides a qualitative summary of key features of state-of-the-art systems contrasted with our proposed method.

The idea of packing multiple low-precision values onto the wider DSP datapath was presented in a Xilinx\tm{} whitepaper~\cite{fu2016deep}. It demonstrates how an average of $1.75$ \texttt{INT8} multiply-accumulate (MAC) operations per clock cycle can be attained per DSP slice. This \texttt{INT8} case was also examined by Lee et~al.~\cite{lee2018double} who even achieved two MAC operations per DSP and cycle. While these techniques increase the achieved operational density specifically for \texttt{INT8} data, they do not try to optimize for other precisions.
Mert et~al. demonstrated how to pack multiple constant multiplications into one DSP~\cite{mert2018efficient}. While this served applications in video processing and compression, it lacks the flexibility of variable inputs. In general, this type of mapping strategy is employed for matrix–vector multiplication (MatVec), which can also be used for convolutions (Conv). However, to date, no work has optimized the operational density for arbitrary precisions in the context of MatVec.

Several works have explored the opportunity to extend the packing to the second multiplier input. Another Xilinx whitepaper~\cite{xilinx2020convolutional} implements this strategy by packing two \texttt{INT4} values into each input datapath of a DSP48E2, thereby, enabling four parallel, mutually independent \texttt{INT4} MAC operations per DSP slice. While Sommer et~al.~\cite{sommer2022dsp} explore overpacking for a further increase of the operational density at the cost of producing approximate results, Mohabbati et~al.~\cite{mohabbati2025dlpack} aim at a further reduction of the data precision describing the packing of unsigned \texttt{INT2} values. Maintaining faithful results across arbitrary precisions, HiKonv~\cite{liu2022hikonv} rather aims at optimizations enabled by the concrete AI application domain specifically requiring the computation of convolutions. Ultimately, they re-discover the technique termed \emph{binary segmentation} by Pan~\cite{pan1993}. It allows and, in fact, exploits the stacking of multiple vertical additions of partial products within the multiplier matrix. Building on these principles, SA4~\cite{yang2024sa4} further advanced the state-of-the-art with an optimized systolic array structure specifically for \texttt{INT4} precision.


A major difference between the individual works lies in how multiple \emph{signed} values are packed into the same wide input path. For two values, this can be done trivially by leveraging the pre-adder found in modern DSP architectures as demonstrated by Fu et~al.~\cite{fu2016deep}. External fabric is typically used for packing more values~\cite{liu2022hikonv, liu2023ssimd}. In contrast to these approaches, we present a method accomplishing this packing, only relying on the pre-adder without the need for additional external resources.

It should be noted that, although similar packing techniques can be used for Altera (Intel) FPGAs, their differing DSP architectures (e.g.,~\cite{Intel:Agilex5VPDSP}) featuring more native support for low-precision operations require other optimization strategies that go beyond the scope of this paper.
\section{Implementation} \label{sec:implementation}

\subsection{Packing Concept} \label{subsec:packing}

\begin{figure}
\centerline{\includegraphics[width=0.5\textwidth]{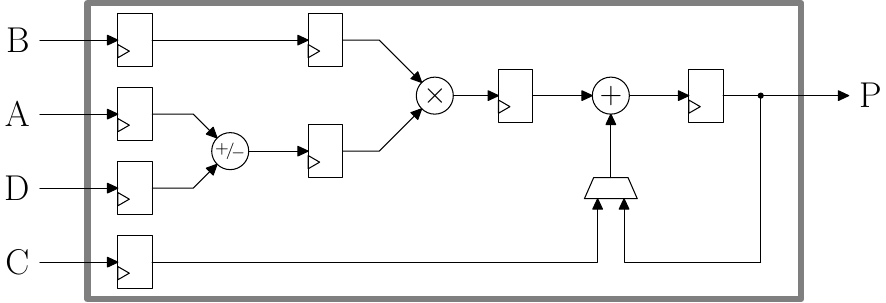}}
\caption{Block-level architecture of the DSP slices.}
\label{fig:dsp_architecture}
\end{figure}

\begin{figure*}
    \includegraphics[width=\textwidth]{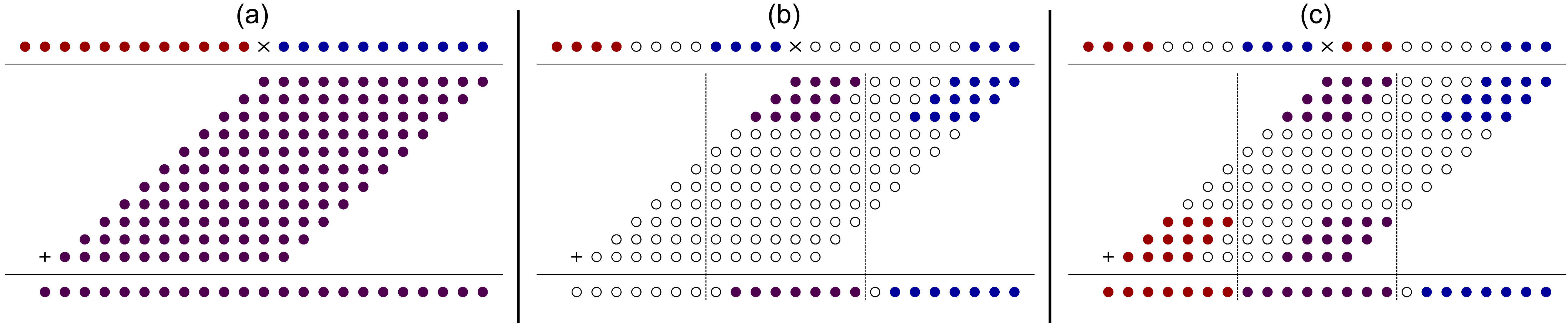}
    \caption{Multiplier utilization options: (a) full-word multiplication, (b) \sdv{} and (c) \bseg{}.}
    \label{fig:circles}
\end{figure*}
\begin{figure*}
\includegraphics[width=\linewidth]{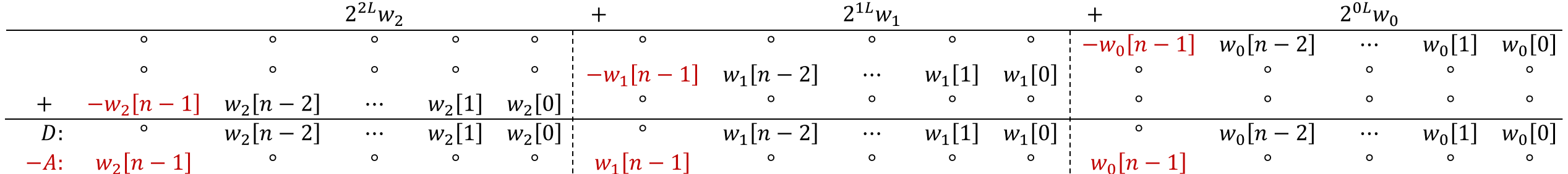}
\caption{Using the DSP pre-adder for packing signed weights by subtracting the separated sign bits.}
\label{fig:sign_packing}
\end{figure*}


We first introduce the packing concept, essential for understanding our proposed architectures. Consider a DSP slice designed with a fixed-width high-precision datapath, as shown in~\fref{fig:dsp_architecture}. It can immediately implement a multiply-accumulate operation with fixed-width input pairs (e.g., $27\times18$-bit for the \texttt{DSP48E2}). However, in many application domains, of which AI is a prominent example, considerably narrower datatypes are common. The naive one-to-one binding of their low-precision compute to DSP slices implies considerable fragmentation cost in unused silicon expenditure. Packing aims at mitigating this price by increasing the number of operations mapped to a single DSP slice.

From now on, we will refer to the technique of applying packing to only one multiplier input as \emph{soft datapath vectorization} (\sdv). In contrast, we call it \emph{binary segmentation} (\bseg) when packing is used on both multiplier input paths. The relevant scope of techniques is illustrated in~\fref{fig:circles}.

Packing must space the inputs sufficiently to ensure that the result lanes can be separated easily. Determining the minimum required distance is essential for achieving the maximum operational density and is discussed in detail for both architectures. Let the lane size $L$, be the number of bits occupied by one value plus the padding introduced to assist the ultimate extraction of individual results. \sdv{} with an embedding of $n$ elements $a_0, \cdots, a_{n-1}$ as packed multiplicand and a shared multiplier $b$ then computes this sum of products:
\begin{align}
\label{eq:product_SDV}
    \left(\sum_{i=0}^{n-1} 2^{i L} a_i\right) \cdot b
    = \sum_{i=0}^{n-1} 2^{i L} \left(a_i \cdot b \right)
\end{align}
If the lane size $L$ is large enough, the high-precision result can be sliced into individual results, thus, enabling the computation of multiple products in parallel.
In contrast, \bseg{} embeds $m$ elements $b_0, \cdots, b_{m-1}$ as packed multiplier instead of a singular $b$, such that it computes:
\begin{align}
\label{eq:product_BS}
    \left(\sum_{i=0}^{n-1} 2^{i L} a_i\right) \! \left(\sum_{j=0}^{m-1} 2^{j L} b_j \right)
    \!=\! \sum_{k=0}^{n+m-2} \! \sum_{\substack{i + j = k \\ 0 \le i < n \\ 0 \le j < m}} \! 2^{kL}a_i b_j
\end{align}
Again, if the lane size $L$ is sufficiently large, the individual results can be extracted from the high-precision output of the multiplier. It is important to note that with \bseg{} some products are added up directly during multiplication, as shown in~\fref{fig:circles} (c), which is exploited for convolutions.

However, the two's complement representation of signed numbers inevitably leads to interference between individual values, even for large lane sizes. The sign extensions of negative lane values overlap all lanes located to their left. This is true for both the vectorized inputs as well as for vectorized outputs. Addressing these interferences is crucial for obtaining correct results.

\subsection{Efficient Packing of Signed Inputs} \label{subsec:negative}
When packing multiple values into a single input datapath, the sign extensions of the two's-complement representation of lower-order values overlap with the lanes of higher-order values. Therefore, plain concatenation no longer suffices if the individual values can be negative. The assumed arithmetic sum must actually be computed. This is conventionally done by \emph{external} additions in the fabric as described by Liu et~al.~\cite{liu2022hikonv}.

However, the packing problem can also be solved directly by leveraging the dedicated pre-adder in the DSP slice architecture of modern AMD FPGAs shown in~\fref{fig:dsp_architecture}. In two's complement arithmetic, the sign bit of a number carries a negative radix weight. After slicing it off from all individual input values, the remainders of their representations can be simply concatenated again into one word $D$. By collecting all the negative-weight sign bits into a second word $A$, the overall packing of an \emph{arbitrary} number of input values can now be concluded by a \emph{single} subtraction of this latter word $A$ from the former word $D$. This method is illustrated in~\fref{fig:sign_packing}. Hereby, the \emph{internal} pre-adder of the DSP slice is configured to compute this subtraction $D-A$.

\begin{figure*}
    \includegraphics[width=\textwidth]{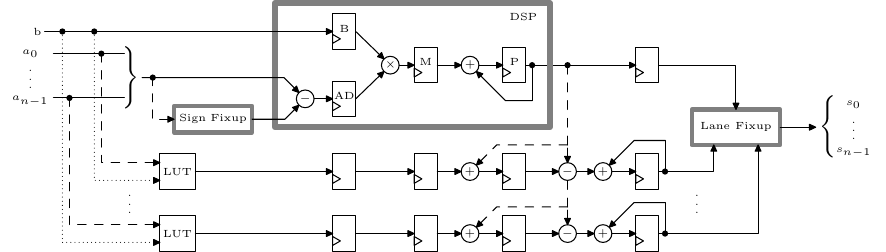}
    \caption{Overall \sdv\ architecture. Dashed and dotted lines indicate the extraction of individual bits.}
    \label{fig:soft_architecture}
\end{figure*}
\subsection{Soft Datapath Vectorization} \label{subsec:soft}
The \sdv{} technique targets matrix-vector multiplications by packing multiple low-precision values into exactly one of the multiplier input words. We now detail the efficient monitoring and correction of the interferences between the compute lanes and what lane sizes are enabled by the implemented technique.

Modern AMD FPGAs feature fracturable LUTs that can implement either one arbitrary Boolean 6-input function or two arbitrary 5-input functions with shared inputs~\cite{Xilinx:UltraScaleCLB}. In the fractured setup, a single LUT is able to compute the two least significant bits of a product given the two least significant bits of both factors. This result allows the computation of a reference for the product accumulation step in terms of the modulo-4 remainder classes of all lanes under the assumption that they were separate. Differences observed in the accumulation results computed by the DSP represent spill-overs between lanes. We will dimension lane sizes such that the arithmetically possible spill-overs are confined to a narrow numeric range of no more than four values. Under this condition, the modulo-4 reference compute enables faithful tracking of all lane spill-overs in an external accumulator. The tracked spill-over totals serve to complete and correct the ultimately emitted lane results. The overall architecture is shown in~\fref{fig:soft_architecture}.
Denote the total spill-over out of lane $i$ as $S_i$ and the result computed for this lane by the DSP as $R_i$. The ultimate result for this lane, $\hat{R}_i$, is the concatenation of $S_i$ as its high part and $R_i$ as its low part, which is reduced by $S_{i-1}$, the spill-over contribution received from its right neighbor lane:
\begin{align}
    \hat{R}_i = \left(2^L\cdot S_i + R_i\right) - S_{i-1}
\end{align}
The proposed packing approach can differentiate four different remainder classes of the spill-overs between accumulation lanes. In the first step, this would allow a lane that is two bits narrower than the computed product as the two topmost bits of this product, which indeed extend into the next lane, can still be recovered and tracked. In a continuous operation, however, the arbitrary value already accumulated in a lane will regularly cause an \emph{extra} carry across its boundary by the addition with the contained part of the product. When accumulating unsigned products, this blurs the possible range of spill-overs from $[0:3]$ to $[0:4]$, in the signed case, from $[-2:1]$ to $[-2:2]$. In either case, the boundaries of the extended ranges can no longer be differentiated by their modulo-4 remainder classes. However, it is sufficient to require a lane size that is, at most, one bit shorter than the width of the accumulated products. Their additions with the already resident accumulated values may still produce an extra carry out of the lane. With only one bit of the product reaching beyond the core lane, this only extends the unsigned spill-over range from $[0:1]$ to $[0:2]$, and the signed one from $[-1:0]$ to $[-1:1]$. In both cases, all possible values are fully differentiated by their modulo-4 remainder class and hence, can be detected and tracked. Denoting the width of the individual factors as $w_a$ and $w_b$, this argument yields the lane size requirement:
\begin{align}
    L \ge w_a + w_b - 1
\end{align}

\begin{figure}[t]
  \centering
  \null\hfill
  \subfloat[\sdv{} on DSP48E2 / DSP58.]{
    \makebox[0.45\columnwidth][c]{
      \includegraphics[width=0.45\columnwidth]{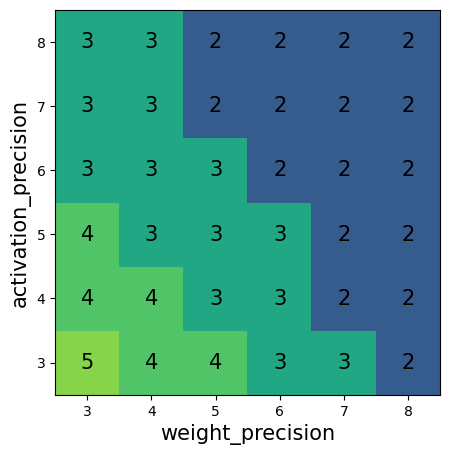}}
    \label{fig:op_density_DOTP}}
  \hfill
  \subfloat[\bseg{} on DSP48E2.]{
    \makebox[0.45\columnwidth][c]{
      \includegraphics[width=0.45\columnwidth]{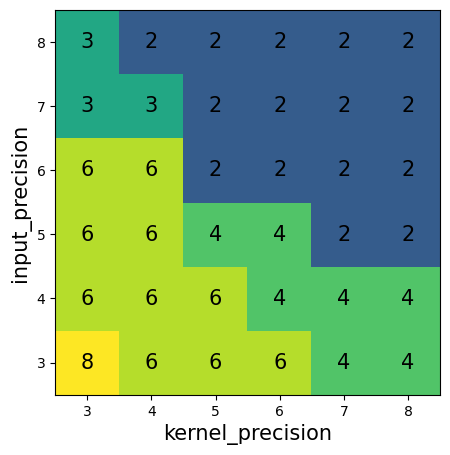}}
    \label{fig:op_density_BS}}
  \caption{Operational density of the proposed methods.}
  \label{fig:op_density}
\end{figure}

This allows the computation of the maximum operational density as a function of the input precisions. The results specifically for the \texttt{DSP48E2} and \texttt{DSP58} slice generations are shown in~\fref{fig:op_density_DOTP}.
These calculations take into account that the leftmost element in a packing does not require a full lane but only its own width plus one bit. The latter is needed to protect its sign. If the elements are unsigned, a leading zero bit ensures correct computation within the underlying signed multiplier. If the elements are signed, the packing with neighboring negative values produces the arithmetic equivalent of a decrement. A padding MSB ensures that this does not flip the overall sign when such a decrement meets the most negative value of $-2^{w_b-1}$ in the leftmost lane.

Note that the \texttt{DSP58} on Versal\tm{} devices supports a native \texttt{INT8} mode capable of accumulating three \mbox{$9\times 8$-bit} products. On this architecture, \sdv\ only adds value for very low-precision computations where it is able to attain an operational density of at least 4~MAC operations per DSP slice.

\subsection{Binary Segmentation} \label{subsec:binary}
We employ the original binary segmentation (\bseg) technique~\cite{pan1993} with non-overlapping computations for implementing the two-dimensional packing of operands on both multiplier input datapaths as shown in~\fref{fig:circles}(c). This approach enables the quadratic increase in operational density with decreasing precision~\cite{liu2022hikonv}. Unlike \sdv, which accumulates all products individually, this \bseg\ implementation achieves the summation of some products directly inside the multiplier matrix. This construction presumes a specific input sharing. While this limits its general applicability, it coincides seamlessly with the compute structure of convolutions.

From an implementation perspective, discrete convolution and correlation are equivalent. One is transformed into the other simply by flipping the kernel. For simplicity, we adopt the correlation formulation, which is commonly used to describe convolutions in deep learning. The 1D convolution of a kernel $K = \left[K[0], \cdots, K[n-1]\right]$ and an input $I=\left[I[0],\cdots,I[m-1]\right]$ with $m\ge n$ elements produces an output of $(m-n+1)$ elements as given by~\eref{eq:conv_1d}.
\begin{align}
\label{eq:conv_1d}
    (K \ast I)[j] = \sum_{k=0}^{n-1} K[k] \cdot I[j+k]\quad\mbox{with }0\le j\le m-n
\end{align}

Particularly in the vision context, the typical convolution is two-dimensional, operating on image frames with a third channel dimension. Multiple channels of the output feature map are computed using independent designated kernels, also known as filters. With $C$ output channels, the set of kernels is a tensor $K \in \R^{W_K \times H_K \times D \times C}$ where each kernel has width $W_K$, height $H_K$ and depth $D$. Such a kernel induces a 2D convolution with an input feature map $I \in \R^{W_I \times H_I \times D}$ with width $W_I \ge W_K$, height $H_I \ge H_K$ and the same depth $D$. The result is a 3D tensor $(K \ast I) \in \R^{(W_I-W_K+1) \times (H_I-H_K+1) \times C}$ with the individual elements given by~\eref{eq:conv_3d}.
\begin{align}
&(K \ast I)[w][h][c] = \nonumber\\
&\sum_{k_w=0}^{W_K-1} \sum_{k_h=0}^{H_K-1} \sum_{d=0}^{D-1} 
K[k_w][k_h][d][c] \cdot I[w + k_w][h + k_h][d] \label{eq:conv_3d}
\end{align}

\begin{figure}
\centerline{\includegraphics[width=0.5\textwidth]{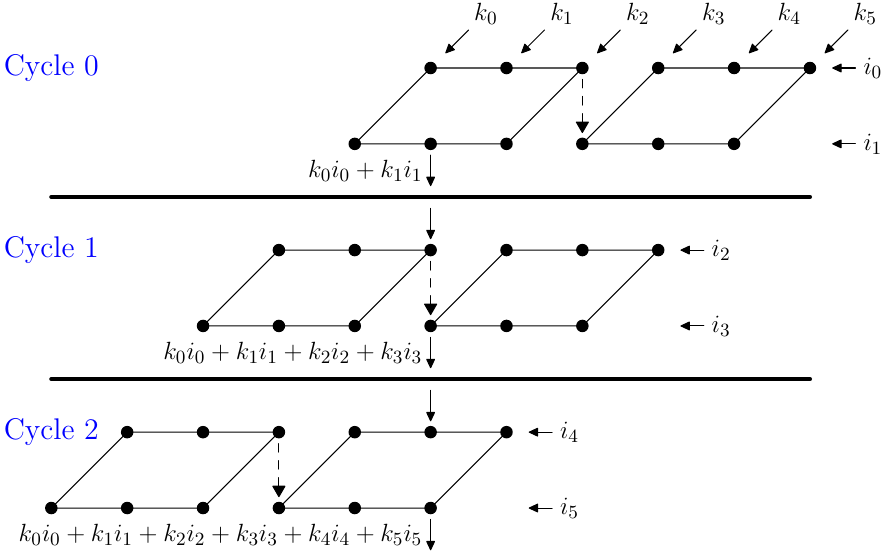}}
\caption{Cycle-by-cycle illustration showing the operation of our \bseg{} architecture.}
\label{fig:BS_Architecture_Basic}
\end{figure}

We will first present the 1D architecture before describing its generalization.
Assuming the packing of $n_k$ kernel elements into the first factor and $n_i$ input elements into the second factor, the multiplier computes all pairwise products between kernel and input elements and adds specific partial results. In each cycle, we want to use all kernel elements and a subset of the input elements. Therefore, for a total of $n$ kernel elements, we require $\lceil n / n_k \rceil$ multipliers. After each cycle, we retrieve the partial results from each lane and shift them to the next set of kernel elements. The situation is illustrated in~\fref{fig:BS_Architecture_Basic} for $n=6,\: n_k=3,\: n_i=2$.
To add the partial results from the previous cycle to the new products, we use the C-input of the DSPs depicted in~\fref{fig:dsp_architecture}. This can be done independently for all lanes by setting only the corresponding bits of the C-input. To minimize the critical path, the connections between DSPs are pipelined, which requires additional buffering of kernel and input elements for synchronization reasons. While the kernel elements must be buffered externally, the input elements can be buffered directly inside the DSPs utilizing the dedicated cascade path of the B-input.

\begin{figure}
\centerline{\includegraphics[width=0.7\linewidth]{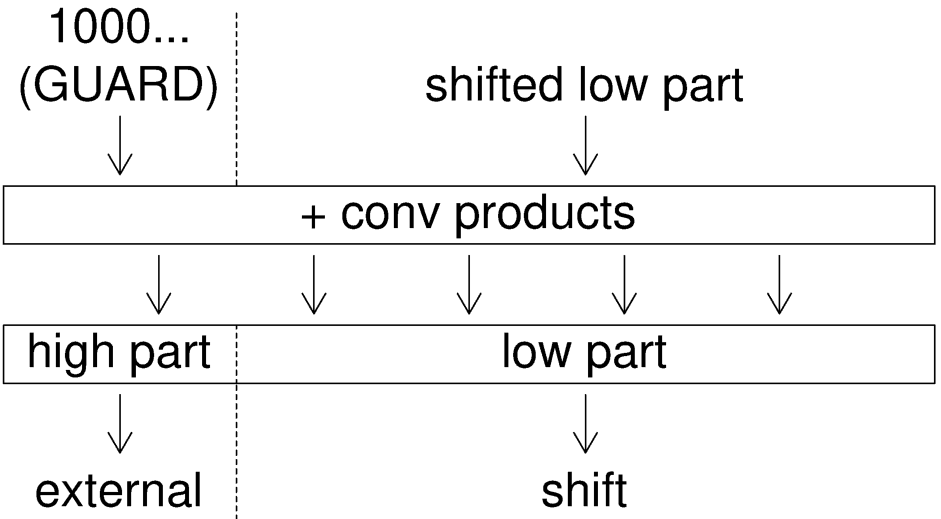}}
\caption{Lane value slicing for multi-stage accumulation.}
\label{fig:Lane_Layout}
\end{figure}

Due to the higher operational density compared to \sdv, the overflow detection technique using the two least significant bits gets more expensive for \bseg. To prevent interference between the individual lanes entirely, every lane is primed with a static offset that biases the accumulation within each lane. These guard bits added to each lane are sufficient to prevent both positive \emph{and} negative spill-overs between lanes. The required offsets can be injected via the C port. In the newer \texttt{DSP48E2} and \texttt{DSP58} slices, they can alternatively be introduced via the \emph{internally} configured \texttt{RND} parameter. The provisioning of guard bits consumes valuable area in the multiplier matrix. We limit this investment to cope only with the accumulation stacked within the height of the multiplier matrix. For the continued accumulation as needed for larger kernels, the lane values are sliced in between DSP stages as shown in~\fref{fig:Lane_Layout}. While the low part stays on the DSP datapath, the high part is extracted and tracked in fabric logic. It is replaced by a guard value that re-biases the lane value, preparing it for the next accumulation stage (cf.~\cite{preusser2020extraction}).

We derive the maximum operational density achievable by this design. The packing of $n_k$ kernel elements of width $w_k$ into the first factor of width $w_A$, and of $n_i$ input elements of width $w_i$ into the second factor of width $w_B$ must satisfy the constraints~~\eref{eq:datapath_a} and~~\eref{eq:datapath_b}.
Similar to \sdv{}, the plus one is needed to protect the sign of the leftmost element.

\begin{align}
    (n_k-1) \cdot L + w_k + 1 &\le w_A \label{eq:datapath_a}\\
    (n_i-1) \cdot L + w_i + 1 &\le w_B \label{eq:datapath_b}
\end{align}

Guard bits separate the individual lanes. To prevent both positive and negative overflows, the offset is chosen as $2^{L-1}$, which centers the accumulation range. This leads to the conditions for avoiding positive overflows, given in~\eref{eq:underflow}, as well as for avoiding negative overflows, given in~\eref{eq:overflow}. These conditions are valid specifically for the case of signed kernel values, unsigned input values, and a low-part width of $w_l$.
\begin{align}
    &2^{L-1} \ge \min(n_k,n_i) 2^{w_k-1} (2^{w_i} \!-\! 1) \label{eq:underflow} \\
    &2^{L-1} > \min(n_k,n_i) (2^{w_k-1} \!-\! 1) (2^{w_i} \!-\! 1) + (2^{w_l} \!-\! 1) \label{eq:overflow}
\end{align}
To determine the maximum operational density $n_k \cdot n_i$, we set $w_l=0$ and drop~\eref{eq:overflow} as it is a weaker condition than~\eref{eq:underflow} in this case. The maximum operational density can, thus, be computed by solving~\eref{eq:datapath_a},~\eref{eq:datapath_b} and~\eref{eq:underflow}. The results are shown in~\fref{fig:op_density_BS}. It can be seen that significantly higher operational densities can be achieved for very low-precision values. Also, the minimum lane size and maximum low part width can be determined from these conditions. However, it may be beneficial to pick a larger lane size if it does not affect the operational density and reduces the width of the external high part. To determine optimal parameters, resource usage is estimated for both the minimum lane size and the minimum lane size plus one. The option with lower overall cost is selected.

The 1D architecture generalizes easily. If the convolution has multiple output channels, these can be computed either in parallel instances or inside the same DSP in a round-robin fashion by buffering partial results instead of advancing to the next kernel elements immediately. Higher-dimensional convolutions are sliced into individual 1D computations whose results are combined before output. In a parallel computation of the rows, an adder tree is used to combine all the results. In a sequential computation, the results are combined using an accumulator.

\subsection{FINN Integration} \label{subsec:finn}
To make our implementations accessible for the wider research community, we have integrated them into AMD's open-source FINN framework~\cite{umuroglu2017finn, blott2018finn}. The SystemVerilog code is available in the supplementary materials and planned to be upstreamed into the official FINN repo~\cite{finn_repo}. This ensures that it can be easily applied to real-world DNN inference tasks.

To ensure compatibility with FINN, our implementations were adapted to its operator interface requirements. This implies data input and output via AXI-Streams assuming a channels-last tensor layout. We adopted and directly optimized these requirements. Consequently, the \bseg{} architecture requires a preceding input generator to not only buffer the stream but to reorder and package the specific elements needed for the parallel DSP inputs. The memory required for this input generator can be implemented using either BRAM or LUTRAM. FINN produces highly customizable dataflow inference solutions. Gauging the trade-off between resource investment and compute throughput is one of its key customization levers, which typically translates into tuning the degree of compute parallelism with the implementation of an operator. Our \sdv{} architecture fits FINN's standard parallelization scheme where \texttt{PE} captures parallelism in terms of outputs and \texttt{SIMD} in terms of inputs. While the \bseg\ implementation escapes this terminology, its parallelism is, nonetheless, flexibly tunable by unrolling along the input width, kernel height and output channel dimensions independently.

\section{Evaluation} \label{sec:evaluation}

A qualitative comparison of the key features between our approach and previous ones has already been shown in~\sref{sec:related} in~\tref{tab:features}. As a quantitative demonstration of the efficiency of the proposed \sdv{} and \bseg{} architectures, we performed a comprehensive analysis of operational density, resource utilization and critical path timing. All results were obtained using out-of-context synthesis in Vivado\tm{} 2025.2 with the default implementation strategy. We targeted the AMD ZCU104 evaluation board, incorporating a Zynq\tm{} UltraScale+\tm{} MPSoC with \texttt{DSP48E2} slices. For clarity, we adopt the following convention: Tensor dimensions are denoted as $\mbox{height}\times\mbox{width}$ for 2D tensors and as $\mbox{height}\times\mbox{width}\times\mbox{depth}$ for 3D tensors. In terms of resource utilization, we will focus primarily on DSP and LUT consumption as the observed results for FF usage are always very similar to LUT consumption. The following evaluation is structured to address three main aspects for each architecture: scalability with respect to precisions and tensor sizes, comparison with state-of-the-art systems, and comparison with previous FINN implementations.

\subsection{Scalability Analysis} \label{subsec:scalability}
To analyze scalability, we performed a parameter sweep over precision and tensor size. We generally targeted and reached a clock frequency of \SI{250}{MHz} for all configurations. For both \sdv{} and \bseg{} the number of DSPs is as expected proportional to the total number of MAC operations per cycle divided by the operational density of a DSP that has already been shown in~\fref{fig:op_density}. However, the amount of additional resources external to the DSPs is equally important.

\begin{figure}[t]
  \centering
  \null\hfill
  \subfloat[Scaling with precision.]{
    \makebox[0.45\columnwidth][c]{
      \includegraphics[width=0.45\columnwidth]{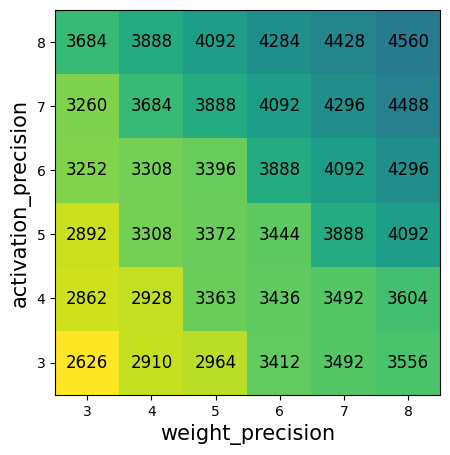}}
    \label{fig:SDV_precision_sweep_LUT}}
  \hfill
  \subfloat[Scaling with matrix size.]{
    \makebox[0.45\columnwidth][c]{
      \includegraphics[width=0.45\columnwidth]{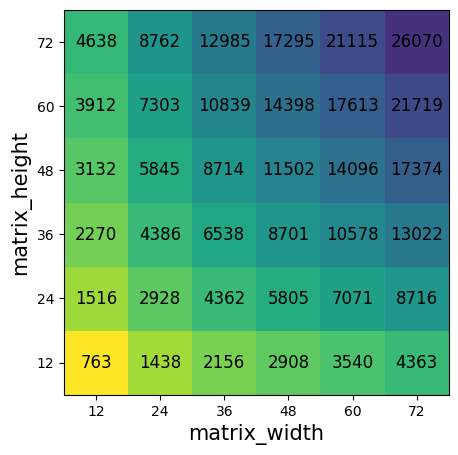}}
    \label{fig:SDV_matrix_sweep_LUT}}
  \caption{Scaling of LUT resource utilization for \sdv{}.}
  \label{fig:SDV_sweep_LUT}
\end{figure}

For the \sdv{} architecture, we use as a reference configuration the multiplication of a $\mbox{24}\times\mbox{24}$-matrix of 4-bit values with a corresponding 4-bit vector, to reflect typical workloads in resource-constrained AI applications. The architecture was configured such that the entire matrix-vector product is computed within three cycles. The LUT resources for the parameter sweep over precision and matrix size are shown in~\fref{fig:SDV_sweep_LUT}.
It is evident that the number of LUTs strongly correlates with the number of DSPs. This implies that a higher packing density is always desirable. Due to the condition that the entire matrix-vector product is calculated in three cycles, a larger matrix also requires more resources. It can be seen that the number of LUTs is approximately a linear function of the total matrix size.

\begin{figure}[t]
  \centering
  \null\hfill
  \subfloat[Scaling with precision.]{
    \makebox[0.45\columnwidth][c]{
      \includegraphics[width=0.45\columnwidth]{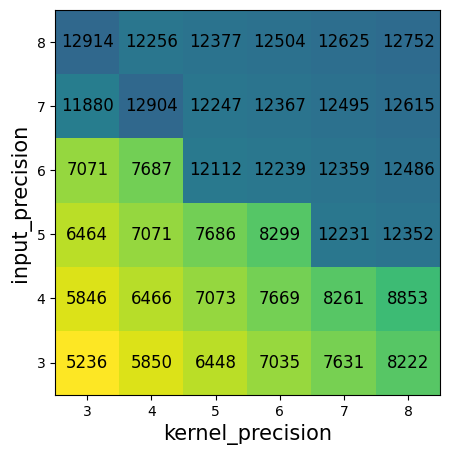}}
    \label{fig:BS_precision_sweep_lut}}
  \hfill
  \subfloat[Scaling with kernel size.]{
    \makebox[0.45\columnwidth][c]{
      \includegraphics[width=0.45\columnwidth]{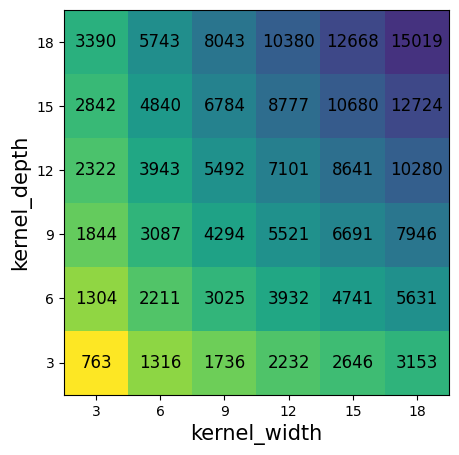}}
    \label{fig:BS_kernel_sweep_lut}}
  \caption{Scaling of LUT resource utilization for \bseg{}.}
  \label{fig:BS_sweep_lut}
\end{figure}

For the \bseg{} architecture, we establish a reference configuration consisting of a convolutional layer with an input feature map size of $\mbox{1}\times\mbox{1500}\times\mbox{16}$ and $128$ kernels of size $\mbox{1}\times\mbox{8}\times\mbox{16}$, both using 4-bit values. This type of configuration is commonly found in audio processing and time-series forecasting, where long temporal sequences are processed with significant channel expansion. Building on this baseline, we perform a parameter sweep over different precisions and kernel sizes, while maintaining an average throughput of eight output feature map elements per cycle. The corresponding LUT resource utilization for these variations is shown in~\fref{fig:BS_sweep_lut}.

Again, the number of LUTs strongly correlates with the number of DSPs, making higher packing densities desirable. Regarding kernel size, the number of LUTs is approximately a linear function of the total kernel size. In addition, there is a constant offset arising from the fixed control logic, as well as a logarithmic component from the increase in output bit width. However, the contribution of the logarithmic component remains marginal across the evaluated range.


\begin{table}[t]
\renewcommand{\arraystretch}{1.2}
\caption{Comparison of the \bseg{} architecture using LUTRAM/BRAM vs state-of-the-art systems for the full UltraNet model (FM) and convolutional layers only (Conv).}
\label{tab:results_ultranet}
\centering
{\setlength{\tabcolsep}{4pt}
\begin{tabular}{l@{\quad}r<{\;}r<{\;}r<{\;}r<{\;}r<{\;}}
\toprule
\textbf{Design} 
& \multicolumn{1}{c}{\textbf{Base}~\cite{ultranet}}
& \multicolumn{1}{c}{\textbf{HiKonv}~\cite{liu2022hikonv}}
& \multicolumn{1}{c}{\textbf{FINN}}
& \multicolumn{2}{c}{\textbf{BSEG}} \\
\cmidrule(lr){5-6}
\textbf{Scope}
& \multicolumn{1}{c}{\textbf{Conv}}
& \multicolumn{1}{c}{\textbf{Conv}}
& \multicolumn{1}{c}{\textbf{FM}}
& \multicolumn{1}{c}{\textbf{FM}}
& \multicolumn{1}{c}{\textbf{Conv}} \\
\midrule
LUT           & 43k        & 48k        & 63k   & 50k/46k & 35k/31k   \\
DSP           & 360        & 327        & 586   & 422 & 422   \\
FPS           & 248        & 401        & 636   & 636 & 636   \\
FPS/DSP       & 0.7        & 1.2        & 1.1  & 1.5 & 1.5  \\
\bottomrule
\end{tabular}
}
\end{table}

\subsection{Comparison with State-of-the-Art Systems}
For the \sdv{} technique, the current state-of-the-art system is represented by~\cite{lee2018double}. It achieves an operational density of two MAC operations per DSP for INT8 values. As depicted in~\fref{fig:op_density_DOTP}, our architecture matches this performance for \texttt{INT8} values. However, as our design supports arbitrary precisions, the achieved operational density significantly increases for many cases below 8 bits. This ensures that the available resources are always fully utilized for aggressive quantization, providing a substantial throughput gain over existing fixed-precision implementations.

For the \bseg{} technique, the most capable state-of-the-art system that also supports arbitrary precisions is HiKonv~\cite{liu2022hikonv}. For comparison, we generated an architecture for the \texttt{INT4} UltraNet model~\cite{ultranet}. Distinct from the original $\mbox{160}\times\mbox{320}$ input resolution~\cite{ultranet}, we employed a square $\mbox{416}\times\mbox{416}$ configuration to align with standard YOLO topologies and to assess the architecture under increased requirements. A comparison of resource utilization and performance of different designs is shown in~\tref{tab:results_ultranet}. In a first step, the entire model was mapped onto the FPGA using FINN. In a subsequent step the first three convolutional layers were replaced by the \bseg{} architecture. Since the reference works~\cite{ultranet, liu2022hikonv} targeted the resource-constrained Ultra96 board and only considered the resources of the convolutional layers, we determined the contribution of these layers as well. Compared to HiKonv~\cite{liu2022hikonv}, our design reduces LUT usage by $27\%$. Although it requires $29\%$ more DSPs to support greater parallelism, it yields a significantly higher throughput, ultimately improving DSP efficiency (FPS/DSP) by $25\%$.


\begin{table}[t]
\scriptsize 
\renewcommand{\arraystretch}{1.2}
\centering
\caption{Resource utilization comparison for the UltraNet convolutional layers: FINN baseline vs the \bseg{} architecture using BRAM (B1) and LUTRAM (B2) for the input generator.}
\label{tab:ultranet_individual_convs}
{\setlength{\tabcolsep}{3pt}
\begin{tabular}{c@{\;\;}r@{\;}r@{\;}r@{\quad}r@{\;}r@{\;}r@{\quad}r@{\;}r@{\;}r@{\;}r@{\;}r@{\;}r@{}}
\toprule
& \multicolumn{3}{c}{\textbf{LUT}}
& \multicolumn{3}{c}{\textbf{FF}}
& \multicolumn{3}{c}{\textbf{DSP}}
& \multicolumn{3}{c}{\textbf{BRAM}} \\
\cmidrule(lr){2-4} \cmidrule(lr){5-7} \cmidrule(lr){8-10} \cmidrule(lr){11-13}
\textbf{Layer}
& \multicolumn{1}{c}{\textbf{FINN}}
& \multicolumn{1}{c}{\textbf{B1}}
& \multicolumn{1}{c}{\textbf{B2}}
& \multicolumn{1}{c}{\textbf{FINN}}
& \multicolumn{1}{c}{\textbf{B1}}
& \multicolumn{1}{c}{\textbf{B2}}
& \multicolumn{1}{c}{\textbf{FINN}}
& \multicolumn{1}{c}{\textbf{B1}}
& \multicolumn{1}{c}{\textbf{B2}}
& \multicolumn{1}{c}{\textbf{FINN}}
& \multicolumn{1}{c}{\textbf{B1}}
& \multicolumn{1}{c}{\textbf{B2}} \\
\midrule
0 & 4959 & 1380 & 2231 &  3248  & 1686 & 2348 &  27 & 18 & 18 &  0    &  1.5 & 0 \\
1 & 7028 & 3536 & 5658 &  12551 & 3725 & 5665 &  72 & 48 & 48 &  0    &   6  & 0 \\
2 & 8465 & 4785 & 6261 &  13335 & 4867 & 6151 &  96 & 64 & 64 & 21.5  &   15  & 11 \\
3 & 4417 & 5871 & 7338 &  7580  & 3954 & 5247 & 144 & 64 & 64 &  8    &   15  & 11 \\
4 & 2746 & 5856 & 6623 &  1420  & 3950 & 4725 &  32 & 64 & 64 &  7.5  &   15  & 11 \\
\bottomrule
\end{tabular}
}
\end{table}

\begin{table}[t]
\renewcommand{\arraystretch}{1.3}
\caption{Comparison of the \bseg{} architecture against the FINN baseline at maximum frequencies.}
\label{tab:results_bs_finn}
\centering
\begin{tabular}{l@{}r<{\quad}r<{\quad}}
\toprule
&\multicolumn{1}{c}{\textbf{FINN Baseline}}
&\multicolumn{1}{c}{\textbf{\bseg{}}}\\
\midrule
Frequency & $\SI{580}{MHz}$ & $\SI{590}{MHz}$\\
LUT       & $17761$         & $6505$\\
DSP       & $256$           & $192$\\
\bottomrule
\end{tabular}
\end{table}

\subsection{Comparison with Previous FINN Implementation}
FINN traditionally lowers convolutions into a sequence of an input generator and a matrix-vector multiplication. To keep the comparison aligned with state-of-the-art techniques, we use the new \sdv{} architecture for matrix-vector multiplications in the FINN baseline. A comparison of the resources required for UltraNet is shown in~\tref{tab:results_ultranet}. The \bseg{} architecture requires $21\%$ fewer LUTs and $28\%$ fewer DSPs at the same throughput. A more detailed comparison is shown in~\tref{tab:ultranet_individual_convs}, where the different types of convolutional layers of the UltraNet were synthesized and compared separately. It becomes apparent that \bseg{} is advantageous for the first three layers, whereas \sdv{} is preferable for all other layers. The problem with layer types 3 and 4 is that the required input generator based on FINN's tensor layout gets costly for many input channels.

To compare both architectures at their maximum operating frequency, we use the base convolutional layer from the scalability analysis in~\sref{subsec:scalability}. The results are summarized in~\tref{tab:results_bs_finn}. We see a significant improvement as the \bseg{} architecture requires $63\%$ fewer LUTs and $25\%$ fewer DSPs while even improving the achievable clock frequency slightly.

\section{Conclusion} \label{sec:conclusion}

In this paper, we presented several novel arithmetic packing strategies that significantly improve the utilization of fixed-width DSP slices in modern FPGA devices. A key contribution is the first efficient method that packs multiple signed values into the same DSP datapath using only the DSP internal pre-adder without any external logic. Furthermore, our proposed overflow detection and guard-bit strategies ensure accurate computations with high efficiency, particularly for precisions below 8 bits, achieving higher operational densities than state-of-the-art techniques.
Based on this, we introduced two complementary architectures, \sdv{} for matrix-vector multiplications and \bseg{} for convolutions, which both support arbitrary precisions. This flexibility allows the design to be used efficiently across a wide range of quantization levels.

We integrated the proposed architectures into AMD's open-source framework FINN and directly optimized them for FINN's standard tensor layout. Experimental results for the full UltraNet model show an efficiency increase by 36\% to \SI{1.5}{FPS/DSP} while reducing the LUT count by 21\%. A comparison of the \bseg{} architecture to the FINN implementation at maximum frequencies shows a reduction in LUT and DSP by 63\% and 25\% respectively, while maintaining or improving achievable clock frequencies. 

Future work will investigate dynamic arithmetic packing strategies that adapt to the computational workload in real time. Additionally, we plan to extend the approach to support the mapping of multiple low-precision floating-point values. Furthermore, our packing techniques are readily compatible with clock pumping such that an even higher throughput per DSP can be achieved in the future.

\ifblinded\else
\renewcommand{\thefootnote}{}
\footnotetext{%
AMD, the AMD Arrow logo, UltraScale+, Versal, Virtex, Vivado, Xilinx
and combinations thereof are trademarks of Advanced Micro Devices, Inc.  Other product names used in this publication are for identification purposes only and may be trademarks of their respective companies.
}
\fi

\bibliographystyle{IEEEtran}
\bibliography{arith26}

@inproceedings{umuroglu2017finn,
  title={{FINN}: A framework for fast, scalable binarized neural network inference},
  author={Umuroglu, Yaman and Fraser, Nicholas J and Gambardella, Giulio and Blott, Michaela and Leong, Philip and Jahre, Magnus and Vissers, Kees},
  booktitle={Proceedings of the 2017 ACM/SIGDA international symposium on field-programmable gate arrays},
  pages={65--74}
}

@article{fu2016deep,
  title={Deep learning with {INT8} optimization on {Xilinx} devices},
  author={Fu, Yao and Wu, Ephrem and Sirasao, Ashish and Attia, Sedny and Khan, Kamran and Wittig, Ralph},
  journal={White Paper},
  institution={Xilinx, Inc.},
  number={WP486 (v1.0.1)},
  year={2017},
}

@book{sze2020efficient,
  title     = {Efficient Processing of Deep Neural Networks},
  author    = {Sze, Vivienne and Chen, Yu-Hsin and Yang, Tien-Ju and Emer, Joel S.},
  year      = {2020},
  publisher = {Morgan \& Claypool},
  doi       = {10.2200/S01004ED1V01Y202004CAC050}
}

@article{blott2018finn,
  title={{FINN-R}: An end-to-end deep-learning framework for fast exploration of quantized neural networks},
  author={Blott, Michaela and Preu{\ss}er, Thomas B and Fraser, Nicholas J and Gambardella, Giulio and O’Brien, Kenneth and Umuroglu, Yaman and Leeser, Miriam and Vissers, Kees},
  journal={ACM Transactions on Reconfigurable Technology and Systems (TRETS)},
  volume={11},
  number={3},
  pages={1--23},
  year={2018},
  publisher={ACM New York, NY, USA}
}

@article{lee2018double,
  title={Double {MAC} on a {DSP}: Boosting the performance of convolutional neural networks on {FPGAs}},
  author={Lee, Sugil and Kim, Daewoo and Nguyen, Dong and Lee, Jongeun},
  journal={IEEE Transactions on Computer-Aided Design of Integrated Circuits and Systems},
  volume={38},
  number={5},
  pages={888--897},
  year={2018},
  publisher={IEEE}
}

@inproceedings{mert2018efficient,
  title={Efficient multiple constant multiplication using {DSP} blocks in {FPGA}},
  author={Mert, Ahmet Can and Azgin, Hasan and Kalali, Ercan and Hamzaoglu, Ilker},
  booktitle={2018 28th International Conference on Field Programmable Logic and Applications (FPL)},
  pages={331--334},
  year={2018},
  organization={IEEE}
}

@article{xilinx2020convolutional,
  title={Convolutional neural network with {INT4} optimization on {Xilinx} devices},
  author={{Xilinx, Inc}},
  journal={White Paper},
  number={WP521},
  year={2020}
}

@inproceedings{liu2022hikonv,
  title={{Hikonv}: High throughput quantized convolution with novel bit-wise management and computation},
  author={Liu, Xinheng and Chen, Yao and Ganesh, Prakhar and Pan, Junhao and Xiong, Jinjun and Chen, Deming},
  booktitle={2022 27th Asia and South Pacific Design Automation Conference (ASP-DAC)},
  pages={140--146},
  year={2022},
  organization={IEEE}
}

@inproceedings{yang2024sa4,
  title={{SA4}: A Comprehensive Analysis and Optimization of Systolic Array Architecture for 4-bit Convolutions},
  author={Yang, Geng and Lei, Jie and Fang, Zhenman and Zhang, Jiaqing and Zhang, Junrong and Xie, Weiying and Li, Yunsong},
  booktitle={2024 34th International Conference on Field-Programmable Logic and Applications (FPL)},
  pages={204--212},
  organization={IEEE}
}

@inproceedings{sommer2022dsp,
  title={{DSP}-packing: Squeezing low-precision arithmetic into {FPGA} {DSP} blocks},
  author={Sommer, Jan and {\"O}zkan, M Akif and Keszocze, Oliver and Teich, J{\"u}rgen},
  booktitle={2022 32nd International Conference on Field-Programmable Logic and Applications (FPL)},
  pages={160--166},
  year={2022},
  organization={IEEE}
}

@inproceedings{liu2023ssimd,
  title={{SSiMD}: Supporting six signed multiplications in a {DSP} block for low-precision {CNN} on {FPGAs}},
  author={Liu, Qi and Sun, Mo and Sun, Jie and Lu, Liqiang and Zhao, Jieru and Wang, Zeke},
  booktitle={2023 International Conference on Field Programmable Technology (ICFPT)},
  pages={161--169},
  year={2023},
  organization={IEEE}
}

@article{mohabbati2025dlpack,
  title={{DLPack}: A {DSP}-Based Low-Bitwidth Packing Architecture for Efficient 2-Bit {CNN} Inference on {FPGA}-based Edge Devices},
  author={Mohabbati, Maryam and Beitollahi, Hakem and Kashi, Somayeh},
  journal={Research Square},
  year={2025},
  note={Preprint},
  doi={10.21203/rs.3.rs-7260088/v1}
}

@manual{Xilinx:UltraScaleCLB,
  author       = "{Xilinx, Inc.}",
  title        = "{UltraScale Architecture Configurable Logic Block User Guide (UG574)}",
  howpublished = "\url{https://www.xilinx.com/support/documentation/user_guides/ug574-ultrascale-clb.pdf}",
  note         = "{UG}574 (v1.6)",
  year         = "2025"
}

@article{thompson2020computational,
  title={The computational limits of deep learning},
  author={Thompson, Neil C and Greenewald, Kristjan and Lee, Keeheon and Manso, Gabriel F},
  journal={arXiv},
  eprint={2007.05558},
  year={2020}
}

@article{boutros2025field,
  title={Field-Programmable Gate Array Architecture for Deep Learning: Survey and Future Directions}, 
  author={Boutros, Andrew and Arora, Aman and Betz, Vaughn},
  journal={Proceedings of the IEEE}, 
  year={2025},
  volume={113},
  number={7},
  pages={613-639},
  doi={10.1109/JPROC.2025.3623023}
}

@manual{Intel:Agilex5VPDSP,
  author       = "{Intel Corporation}",
  title        = {Variable Precision DSP Blocks User Guide: Agilex 5 FPGAs and SoCs},
  howpublished = "\url{https://docs.altera.com/r/docs/813968/25.1/variable-precision-dsp-blocks-user-guide-agilextm-5-fpgas-and-socs/agilextm-5-variable-precision-dsp-blocks-overview}",
  note         = "{UG}-813968 (v25.1)",
  year         = "2025"
}

@article{pan1993,
  author = {V. Pan},
  title = {Binary Segmentation for Matrix and Vector Operations},
  journal = {Computers \& Mathematics with Applications},
  volume = 25,
  number = 3,
  pages = {69-71},
  year = 1993,
  issn = {0898-1221},
  doi = {https://doi.org/10.1016/0898-1221(93)90144-K}
}

@patent{preusser2020vectorization,
  title={Vectorization of Wide Integer Data Paths for Parallel Operations with Side-Band Logic Monitoring the Numeric Overflow Between Vector Lanes},
  author={Preusser, Thomas B. and Branca, Thomas A.},
  nationality={United States},
  number={10,671,388 B1},
  year={2020},
  month={June},
  day={2},
  assignee={Xilinx, Inc.}
}

@patent{preusser2020extraction,
  title={Vectorization of Wide Integer Data Paths into Parallel Operations with Value Extraction for Maintaining Valid Guard Bands},
  author={Preusser, Thomas B. and Branca, Thomas A.},
  nationality={United States},
  number={10,747,534 B1},
  year={2020},
  month={Aug},
  day={18},
  assignee={Xilinx, Inc.}
}

@misc{ultranet,
  title={{UltraNet}: A {FPGA}-based Object Detection for the {DAC-SDC} 2020},
  year={2020},
  author={Kang Zhan and Junnan Guo and Bingyan Song and Wenbo Zhang and Zhenshan Bao},
  howpublished = {\url{https://github.com/heheda365/ultra_net}}
}

@misc{finn_repo,
  title={{FINN}: Fast, Scalable Quantized Neural Network Inference on {FPGAs}},
  howpublished = {\url{https://github.com/Xilinx/finn}}
}

\end{document}